\documentclass[]{spie}  

\usepackage{amsmath,amsfonts,amssymb}
\usepackage{graphicx}
\usepackage[colorlinks=true, allcolors=blue]{hyperref}
\usepackage{fancyvrb}

\def\lsim{\mathrel{\lower2.5pt\vbox{\lineskip=0pt\baselineskip=0pt
          \hbox{$<$}\hbox{$\sim$}}}}

\def\gsim{\mathrel{\lower2.5pt\vbox{\lineskip=0pt\baselineskip=0pt
          \hbox{$>$}\hbox{$\sim$}}}}

\title{HEART: automated build and test infrastructure for real-time controller development}

\author[a]{Edward L. Chapin$^*$}
\author[a]{Jennifer Dunn}
\author[a]{Dan Kerley}
\author[a]{Lianne Mueller}
\author[a]{Malcolm Smith}
\author[a]{Jonathan Stocks}

\affil[a]{National Research Council Herzberg, 5071 W Saanich Rd, Victoria, V9E 2E7, Canada}

\authorinfo{$^*$ed.chapin@nrc-cnrc.gc.ca, phone +1 250 363 6971}

\begin{document}
\maketitle

\begin{abstract}
The Herzberg Extensible Adaptive optics Real-Time Toolkit (HEART) is a complete framework written in C and Python for building next-generation adaptive optics (AO) system real-time controllers, with the performance needed for extremely large telescopes. With numerous HEART-based RTCs now in their design or build phases, each with different AO algorithms, target hardware, and observatory requirements, continuous automated builds and tests are a cornerstone of our development effort. In this paper we describe the many levels of testing that we perform, from low-level unit tests of individual functions, to more complex component and system-level tests that verify both numerical correctness and execution performance. Incorporating extensive testing into HEART since its inception has allowed us to continuously (and confidently) refactor and extend it to both meet the changing needs of local on-sky experiments, as well as those of the several major facility instruments that we are developing.
\end{abstract}


%
%

\section{INTRODUCTION}
\label{sec:intro}

The Herzberg Extensible Adaptive optics Real-Time Toolkit (HEART) is a framework being developed at the Herzberg Astronomy and Astrophysics Research Centre (part of National Research Council Canada, NRC-HAA) to construct real-time controllers (RTCs) for a number of AO instruments, including the Research, Experiment and Validation of Adaptive Optics with a Legacy Telescope (REVOLT) on-sky testbed which is currently active on the 1.2-m telescope at the Dominion Astrophysical Observatory (DAO) in Victoria, Canada\cite{dunn2024,jackson2024,vankooten2024a,vankooten2024b}, and several major upcoming facility-class instruments. The central purpose of an RTC is to process data provided by wavefront sensor cameras (WFSs) and to generate commands for wavefront correctors (WCs, such as deformable mirrors and tip/tilt stages) at frame rates typically ranging from ~500--2000\,Hz. This Hard Real-Time (HRT) component has stringent requirements on latency and jitter. Other activities include external command and status reporting services, the collection of statistics and generation of new control parameters, and the storage and analysis of recorded data, which we collectively refer to as Soft Real-Time (SRT) tasks. Our design was originally driven by the challenging requirements of the multi-conjugate laser guide star Narrow Field InfraRed Adaptive Optics System (NFIRAOS) for the Thirty Meter Telescope (TMT). Early in its design phase several approaches were considered, some including specialized hardware such as Field Programmable Gate Arrays (FPGAs) to meet the tight real-time requirements. However, benchmarking experiments by our team demonstrated that a general-purpose CPU-based control system would be adequate\cite{smith2016}, thus simplifying the expected build and long-term maintenance effort since commercial off-the-shelf (COTS) components and common programming languages and software development tools can be used. It was later recognized that the approach for the NFIRAOS RTC could be generalized into a framework of reusable components, and HEART was born. HEART is capable of many flavours of AO systems including single conjugate, multi-conjugate, and multi-object AO, using both modal and zonal reconstructors, multiple wavefront sensors and wavefront correctors \cite{kerley2019}. In addition to REVOLT and NFIRAOS\cite{atwood2024}, HEART-based RTCs are now under construction for the Gemini North Adaptive Optics system (GNAO)\cite{sivo2024}, the Gemini Planet Imager 2.0\cite{chilcote2024}, and the ArmazoNes high Dispersion Echelle Spectrograph (ANDES) for the Extremely Large Telescope \cite{pinna2024}. It is designed to accommodate both single and multi-server systems using base software components such as circular buffers and telemetry to manage the flow of data, and processing ``blocks'' which incorporate streamed triggering mechanisms to facilitate pipelining, whereby the calculations for a frame can begin as soon as the first pixels from a WFS arrive\cite{smith2022}.

We have been developing HEART continuously since 2019 with a team typically numbering $\sim$5-6 active developers. At the time of writing there are approximately $\sim$125\,k lines of C, $\sim$15\,k lines of Python, $\sim$12\,k of Javascript, and $\sim$5\,k of HTML source code\footnote{Measured using pygount, \texttt{https://pypi.org/project/pygount/}}. Our first on-sky demonstrations on REVOLT occurred in 2022\cite{gamroth2022}. Automated testing has been a cornerstone of this effort since its inception, and we have comparable amounts of library/application code and test code. During this initial build phase we have adopted a straightforward continuous build environment whereby any merges of new code into the master branch triggers compilation of the entire code base, the execution of the full test suite on standard hardware, and the generation of documentation including a detailed test report. While this system has served us well, we are approaching a challenging new phase in which we will have to both support previously delivered systems and newer RTCs under development, with a range of differing operational requirements (including different operating systems, hardware constraints and external interfaces). We are thus in the process of updating our infrastructure to better support these broad goals. Section~\ref{sec:overview} gives an overview of our development process and tools. Details of the various types of automated tests that we perform are described in Section~\ref{sec:test}. Finally, Section~\ref{sec:new} addresses updates to our system that will make it easier for us to continue developing and maintaining RTCs for multiple clients.

%
%
\section{OVERVIEW OF DEVELOPMENT PROCESS AND TOOLS}
\label{sec:overview}

Like many modern instrumentation software projects, HEART and RTCs that depend on it are developed by our group using a mixture of agile concepts and tools, with a hybrid process for marrying our work to waterfall designs for the hardware that it controls (instruments, telescopes). Overall software designs are decomposed into large-scale ``epics'' which are typically major pieces of functionality that can be tied directly to design requirements, and placed into our backlog. During the build phases of our projects the team further decomposes epics into smaller ``stories'' that typically add incremental functionality to our software, and which can be built within our fixed-cadence 1-month sprints (i.e., the time to complete a story typically range from a day up to $\sim$1--2 weeks). A story is considered complete when: (i) code changes required to implement the story have been implemented; (ii) tests have been written to demonstrate that the changes work; and (iii) any new functionality is documented. Overall progress is then estimated and reported regularly to stakeholders based on the number and fractions of the epics that have been completed as compared to the original plan. Since our team is regularly balancing development to support multiple projects, stories are scheduled in the upcoming sprint from our backlog based on current or upcoming constraints (e.g., certain functionality may be required at a scheduled time for lab testing, or a final delivery to an observatory is approaching). We also regularly add new stories to the backlog to manage dynamic circumstances such as: (i) refactoring activities; (ii) bug fixes; and (iii) scope changes (new or changed functionality). Periodic backlog ``grooming'' sessions are used to review and re-prioritize / update / discard older stories that have not yet been scheduled.

\begin{figure}[hbt]
    \begin{center}
        \includegraphics[width=0.9\linewidth]{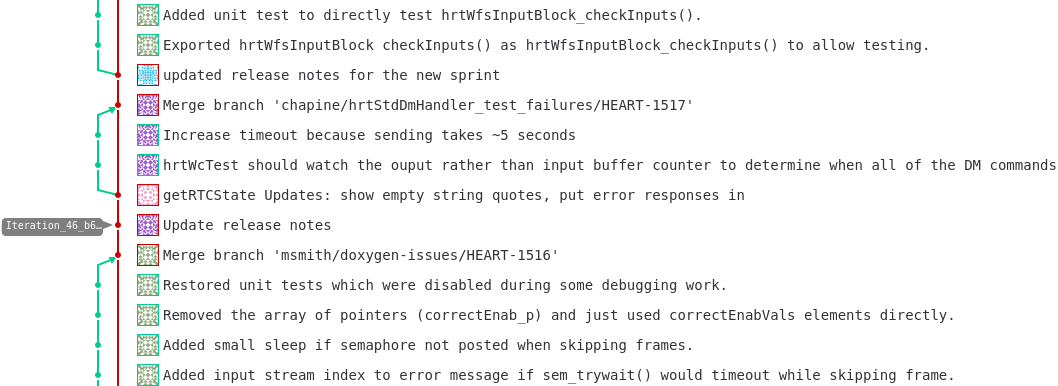}
    \end{center}  
    \caption{\label{fig:branch}
    Example of the semi-linear git repository branching style: story branches are rebased on master and then merged with the \texttt{--no-ff} option. Branch names include the lead developer, a brief description, and a Jira reference, which appear in the merge commit message making it easy to track work. At the end of each month-long sprint a release is made, and the master branch is tagged at that point, e.g., ``Iteration\_46'' as shown above.
    }
\end{figure}

Jira is used to manage the story backlog and sprint planning. All of our code is organized into git repositories, and we use a ``semi-linear'' branching strategy. To summarize: (i) stories are developed in branches off of master in a shared repository with the following naming convention: \texttt{[lead~developer~username] /[brief~description] / [Jira~story~number]}; (ii) when ready, the branch is first rebased on the tip of master (\texttt{git rebase master}), and then (iii) a merge commit to master is forced using \texttt{git~merge~[branch-name]~--no-ff}. The result is that all commits for a story are applied in sequence at the tip of master, but with a merge commit that encodes the informative branch name. This style was adapted from that used for the Rubin Observatory\cite{jenness2018}. While the semi-linear branching approach is less common, it helps with visualization. An example is shown for our main HEART repository in Figure~\ref{fig:branch}. The only downside is that early commits in a branch may not compile and/or pass tests because the original branch point may have been at an earlier point on master. Merge conflicts are dealt with during the rebase, and additional commits at the tip of the branch after rebasing to get tests working again prior to the merge to master are not uncommon. While extra time could be taken to clean up branches prior to merging into master (e.g., using an interactive rebase), we believe we have found a good balance of cost/benefit with the software engineering practices adopted by our small team.

\begin{figure}[hbt]
    \begin{center}
        \includegraphics[width=0.9\linewidth]{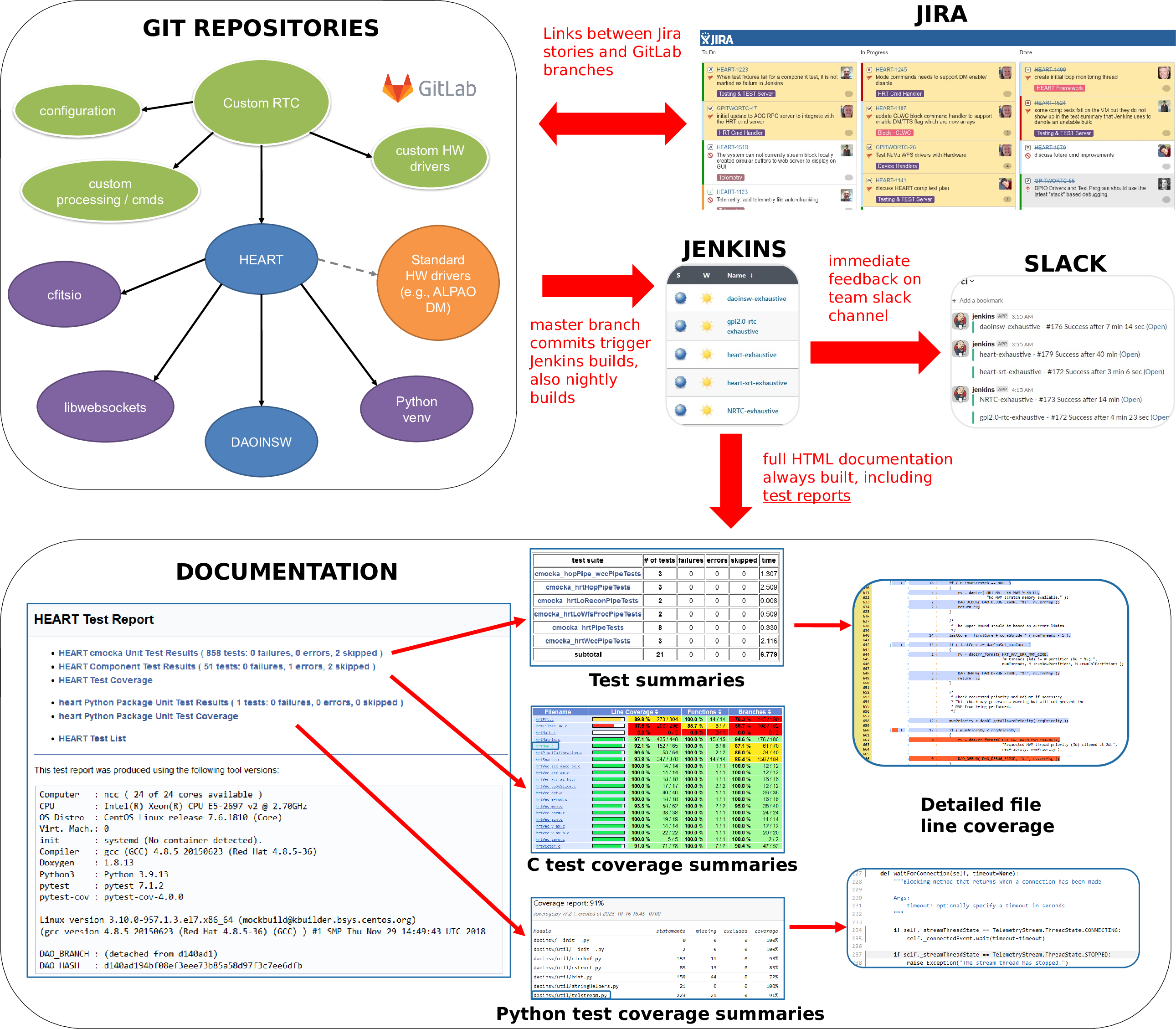}
    \end{center}  
    \caption{\label{fig:overview}
    Current build and test infrastructure. Source code is hosted in a local GitLab instance. A particular RTC will have custom code and configuration in its own repository (green), and uses submodules to manage versions of the main two dependencies: HEART, and a lower-level instrument software library DAOINSW (DAO instrument software) which we develop (blue). Additional dependencies include \textsf{cfitsio}, \textsf{libwebsockets}, and a minimal set of Python packages (e.g., \textsf{NumPy}, \textsf{AstroPy}) managed through a virtual environment (purple), and optional vendor SDKs for standard hardware that HEART supports (orange). Work is organized into stories and planned using Jira. A local instance of Jenkins executes builds whenever there is a commit to master, and as part of a set of nightly builds. Build results are automatically posted to a team slack channel to give immediate feedback. Full documentation, including the detailed test and coverage reports are always built by Jenkins and available to developers as HTML pages.
    }
\end{figure}

All of our git repositories are hosted in a local instance of GitLab-CE\footnote{https://gitlab.com/opensource/gitlab-ce}. Referring to Figure~\ref{fig:overview}, each RTC resides in its own git repository, including custom configuration, software, device interfaces and documentation. They then have two principal dependencies managed through git submodules: the main HEART repository, and a lower-level DAO instrument software (DAOINSW) repository which includes things like system configuration and networking, socket communication, error handling and logging libraries. Our primary external build dependencies are: \textsf{cfitsio} and \textsf{libwebsockets} for C code, and several common Python modules managed through virtual environments (\textsf{NumPy}\footnote{https://numpy.org/}, \textsf{AstroPy}\footnote{https://www.astropy.org/}). Additionally, we use \textsf{Doxygen}\footnote{https://www.doxygen.nl/index.html} for documentation, \textsf{doxypypy} so that Python docstrings can be included in our \textsf{Doxygen} documentation, and \textsf{cmocka} for testing C code.

We use GitLab web hooks to trigger builds in a local Jenkins instance every time there is a commit to master. The dependencies between projects (i.e., an RTC depends on HEART, HEART depends on DAOINSW) are expressed using ``Build other projects'' in the ``Post-build actions''. Build artifacts from upstream processes persist in the Jenkins workspace directories and can be used by the downstream projects. While this strategy works, it does preclude re-building upstream projects while downstream builds are running. Different make targets are used to build all of the code, execute tests, and produce documentation all as part of the same job (including the test and coverage reports). Slack integrations with Jenkins are used to alert the entire team if a build fails. Additionally, we have configured a set of nightly Jenkins builds to help identify transient runtime issues. There are two major shortcomings of this approach. First, tests do not necessarily always pass on the master branch, which is typically caused by differences in a developer's computer environment that may allow tests to pass for them, but not in Jenkins. Second, due to the nature of our infrastructure and test requirements, only a single build and test suite execution can occur at a time on our Jenkins build host, which results in a bottleneck. These and other issues are some of the motivations for the new system that is described in Section~\ref{sec:new}. At the end of our monthly sprints, when all tests are passing, we generate a release tag on the master branch. We then create a single commit on a separate release branch that is a squash of all the commits in the sprint which we provide to our clients. 

The continuously-built \textsf{Doxygen} documentation is extensive, ranging from high-level system descriptions, release notes, stand-alone user guides and FAQs, to detailed low-level in-line documentation of code. \textsf{Doxygen} has excellent support for C and Javascript out-of-the-box, and as noted earlier, we are able to incorporate Python docstrings using \textsf{doxypypy}. The \textsf{Doxygen} syntax is similar to LaTex, and the ability to include Tex-style mathematics is frequently beneficial. The output web pages are immediately available to developers, making it easy to see which tests (if any) have failed, and an overview of the number of lines of code that are covered by the tests. A portion of an example test report for HEART is shown at the bottom of Figure~\ref{fig:overview}.

%
%
\section{Test strategies}
\label{sec:test}

Testing of HEART software is a multi-tiered effort that exercises everything from low-level function calls in libraries, up to the behavior of a full system which involves multiple applications running concurrently. This section covers some of the main aspects and challenges that we have encountered during the initial build phase.

\subsection{C code}
\label{sec:c_tests}

All of the HRT, and some of the SRT portions of HEART are written in C.  We use the \textsf{cmocka}\footnote{https://cmocka.org/} testing framework both for unit tests and some of our component testing. We generally have a single \textsf{cmocka} test file for every C library code file. Executing the \textsf{cmocka} test binaries with \texttt{gcov} allows us to measure code coverage. It is also useful to execute these binaries with \texttt{valgrind} to help identify things like invalid memory access. Many tests simply provide inputs to single functions and verify that they provide expected outputs. However, much of our testing involves significant numbers of nested calls, and verifying that error conditions from things like low-level system routines are handled correctly. Mocking is therefore used extensively in our C test suite, for which our approach has generally been: (i) write a wrapper function that either calls the real function or returns a pre-defined values using global variables; (ii) use the \texttt{-Wl,--wrap=[funcName]} linker arguments to replace the real function with the wrapped function in our test binaries; and (iii) turn mocking on and off as needed throughout the tests using the global variables.

The most complicated tests that we perform with \textsf{cmocka} are for higher-level multi-threaded abstract components. A good example is a HEART processing ``block'', which in broad terms accepts inputs either from circular buffers or commands, performs calculations, and produces outputs, typically sent to circular buffers. A HEART block always has at least two threads: a command handler, which receives commands over a UDP socket, and a worker, which continuously checks for new inputs and performs the calculations. Depending on the complexity of the block, test setup may involve creating many input and output circular buffers, as well as issuing several commands to test different states of the block. It is also sometimes necessary for the test code itself to spawn additional threads if the block is expecting asynchronous inputs. For example, a WFS input block will expect pixel data to arrive over a socket; the test code can use a thread for this purpose. To ensure deterministic behavior, we may use semaphores to trigger each step of the test in these auxillary threads. For example, in the WFS input pixel case, the main test thread might spawn both the pixel-generating test thread and the block, and then have a for loop over pixels; each time through it posts a semaphore (so that the pixel thread sends more data), waits for the block worker thread to respond by polling its outputs, and then verifies that the processing was performed correctly before continuing. One issue we have found with with these multi-threaded \textsf{cmocka} tests is that we have to be very careful with the setup and teardown. Since each test in the suite is intended to be independent of the other tests, the teardown needs to ensure that any extra threads are halted (ideally with a graceful shutdown, but falling back to cancel/kill) -- especially if an earlier test failed and exited prematurely.

As shown in Figure~\ref{fig:unit_tests}, we use in-line \textsf{Doxygen} commands to help describe and organize our test functions to ensure clarity in test reports (Figure~\ref{fig:overview}).

\begin{figure}[hbt]
    \centering
    \scriptsize
    \begin{minipage}[t]{0.48\linewidth}
    \begin{Verbatim}[frame=single]
/*!
 \test \ref  test_hrtTelemetry_badFileSize <br>

 \addtogroup hrtTelemetryTests
 \{
 \copybrief test_hrtTelemetry_badFileSize
 <hr>
 \}

 \brief Tests to detect problems with truncated files
 */
static void test_hrtTelemetry_badFileSize()
{
    daoErr_t rv;
    hrtCB_circBuffer_t *cb1_p = NULL;
    hrtCB_circBuffer_t *cb2_p = NULL;

    DAO_DEBUG( DAO_DEBUG_DIAG, "Test: %s\n", __func__ );

    //
    // Create cb1
    //
    rv = createTestCB( 5, &cb1_p );
    assert_string_equal( rv.errMsg, "" );


    \end{Verbatim}
    \end{minipage}
    \begin{minipage}[t]{0.48\linewidth}
    \begin{Verbatim}[frame=single]
def test_circBuf_BucketSerializer_columnMajor(cleanup_files):
    """
    \test \ref test_circBuf_BucketSerializer_columnMajor

    \addtogroup daoCircBufUnitTests
    @\{
    \copybrief test_circBuf_BucketSerializer_columnMajor
    \}

    \brief Tests serialization round-trip with columnMajor

    """

    specs = cbSpecsStruct.makeDict("testCBColumnMajor",
                                    CbDt.MATRIXFLOAT,
                                    10, 2, 3, 1,
                                    "description")

    bs = BucketSerializer(specs)

    x = np.array([[1, 2, 3], [4, 5, 6]])
    rawData = bs.serialize(Bucket(x))
    newBucket = bs.deserialize(rawData)
    y = newBucket.data

    np.testing.assert_equal(x, y)    
    \end{Verbatim}
    \end{minipage}
    \vspace{0.5cm}
    \caption{\label{fig:unit_tests}
    Example snippets of a \textsf{cmocka} unit test in C (left), and a \textsf{pytest} unit test for Python (right). In each case we use a number of \textsf{Doxygen} commands in the comments to assist with the generation of clear and informative test reports.
    }
\end{figure}

\subsection{Python code}

Python is used for numerous SRT tasks (including calculation of control system parameters using statistics provided by the HRT). Note that we do not presently have any Python wrappers for HEART C code. This choice was made so that all of our Python code can be distributed and used independently, though some communication interfaces required native Python implementations for interoperability with the C code. We use \textsf{pytest} both to test native Python code (and measure coverage), as described in this section, and also as a framework for many of our higher-level component and system-level tests (when multiple processes are involved) -- see Section~\ref{sec:system}. Like our \textsf{cmocka} tests, we use Python docstrings to document tests, and \textsf{doxypypy} allows us to incorporate \textsf{Doxygen} commands in the docstrings, giving our test reports a common look and feel. Since we use \textsf{NumPy} arrays extensively, we frequently make use of the \textsf{NumPy} testing module to make assertions about the calculations of results (Figure~\ref{fig:unit_tests}). In these cases, if an \texttt{Exception} is thrown by an assertion it will provide a report indicating how many of the array elements were mismatches, what and where the differences were, etc. Coverage is measured by executing our tests with \textsf{pytest-cov}, and integrating the results into the full test report.

\subsection{Hardware interfaces}
\label{sec:hw}

To test code that interfaces with hardware -- primarily the WFS input block, and the WC output block -- we have implemented standard UDP-based protocols for sending pixels and receiving wavefront corrector commands. These protocols serve the dual purposes of providing a reference for instrument builders (for example, a custom-built WFS camera could implement this protocol natively, though software development kits are used for most hardware as described below), but also to enable us to develop stand-alone hardware test applications that can be used to feed simulated data into, and read wavefront corrector commands out of HEART test applications (that latter of which can write the commands to a file for later analysis). This approach is particularly suitable for the black-box tests that are described in Section~\ref{sec:system}. Lower-level \textsf{cmocka} tests for individual blocks (Section~\ref{sec:c_tests}) can simply use these protocols directly from within the test process, without requiring stand-alone testers.

Most real hardware interfaces involve third-party software development kids (SDKs). They are incorporated into HEART by wrapping them in standard ``handler'' functions with particular argument signatures and expected input and output data structures. These wrappers are then provided as function pointers to the HEART modules that use them. The HEART \texttt{device} library makes use of build flags to establish which SDKs are available, and to set appropriate compiler and linker arguments. A utility function is provided for returning a requested set of handler functions (as well as detecting and generating an error if support for the requested handler was not built into the library) so that a particular type of device can be selected at runtime through the use of a configuration file (along with any special connection or device-specific configuration information). This feature is particularly useful for switching between real hardware and a stand-alone simulator without recompilation.  The HEART device library supports several standard pieces of hardware out-of-the-box (if the developer is in possession of the required SDKs), and can also be extended by downstream RTCs to support custom hardware required by clients. The use of these hardware handler functions has also made it trivial to create stand-alone laboratory test applications for each hardware type. Our generic WFS and WC testers are simply fed the appropriate hardware handler function pointers when they are instantiated: the former is used to read from a camera and write pixels to a file and/or stream them to another location over a socket, while the latter can read commands from a file and/or receive them over a socket before sending them to a real wavefront corrector.

When vendor SDKs are not available we have implemented a trivial set of mocked SDK function calls so that we are still able to compile all of the hardware handlers in their absence. In the future we will plan to augment these mocked functions so that their return values and side effects can be controlled from test code (thus enabling \textsf{cmocka} testing of the main handler code branches).

\subsection{Black-box component and system testing}
\label{sec:system}

In order to test entire applications, or systems of applications running concurrently (as in a full RTC), we require a method for orchestrating the setup and teardown for all of the processes involved, sending inputs, and making assertions about the outputs. To perform these tasks we have adopted \textsf{pytest} -- partly because we are already using it for Python library code testing -- but also because Python and \textsf{pytest} have a number of features that make it easy to setup, run, debug, and get reports from these tests.

\begin{figure}[hbt]
    \begin{center}
        \includegraphics[width=0.75\linewidth]{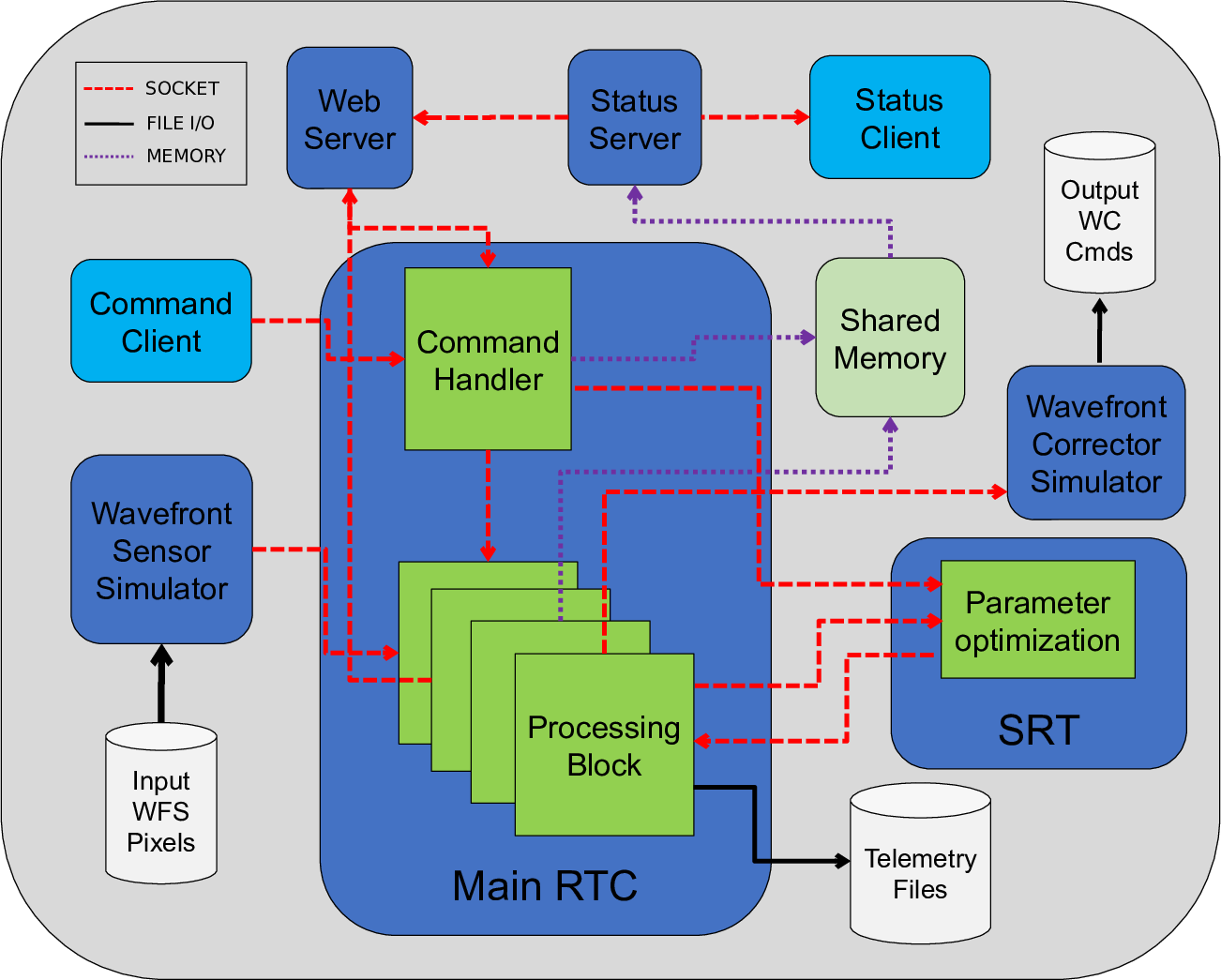}
    \end{center}  
    \caption{\label{fig:system}
    Simplified overview of a full RTC system indicating the main processes that are involved, and methods of communication. The long-running processes are indicated as dark blue boxes, while small client applications are shown in a light blue.
    }
\end{figure}

To get a sense of the complexity of these tests Figure~\ref{fig:system} depicts a simplified view of a running RTC system. The principal long-running processes are depicted in darker blue: the ``Main RTC'' process (primarily the HRT), the ``SRT'' which is a Python application, the ``Wavefront Sensor'' and ``Wavefront Corrector'' simulators (which respectively feed known pixels into the RTC from one file, and write the generated commands to another file), the ``Status Server'' which monitors a Shared Memory structure, and a ``Web Server'' which provides the interface to HEART GUIs. Once running, these processes primarily communicate with each other over network sockets. Commands are provided to the system from the outside world using a TCP protocol (which can be generated either from a stand alone ``Command Client'' application, or the GUI via the Web Server). Commands are passed internally within the Main RTC using a more performant UDP-based protocol (but with message sizes limited to single UDP packets, e.g., $\lsim$8\,kB). Within the Main RTC, data is transmitted between blocks using circular buffers. Data is sent between processes using ``Telemetry Streams'' over network sockets. Telemetry can also be streamed to on-disk files for later analysis. Finally, the Shared Memory structure is written to by the Main RTC, primarily to report status information, and the Web Server or external applications (via the stand-alone ``Status Client'') can query this memory asynchronously.

We use \textsf{pytest} fixtures to simplify the startup and shutdown of the various applications required for a given test. As an example, the following code shows how the main RTC application, ``scaoTemplate'', is often started within a fixture in our tests:

\footnotesize
\begin{Verbatim}[frame=single]
@pytest.fixture
def cmdHandler(cleanup_gms, cleanup_telFiles, dmWooferSimulator, dmTweeterSimulator, ttSimulator):
    """
    Startup and shutdown of \ref scaoTemplate
    """

    # === Setup ===

    proc = None
    configPathFile = DEFAULT_CONFIG_PATH + "/" + DEFAULT_CONFIG_FILE_NAME
    cmdLine = ["scaoTemplate", "-config", configPathFile,
               "-host", "compTestScaoTemplate"] #, "-d", "4"]
    proc = ServiceProc(cmdLine,
                       expect_str="About to create listening address")
    
    yield cmdHandler

    # === Teardown ===

    if proc is not None:
        print("Shutting down cmdHandler")
        proc.stopProc()
\end{Verbatim}
\normalsize

The first thing to note is that fixtures can be nested; \texttt{cmdHandler} depends on several other fixtures which perform activities like cleanup from previous runs, and starting up other applications that this fixture depends on. By expressing the dependency relationships in this way, \textsf{pytest} implicitly starts up everything in the right order. Next, everything up to \texttt{yield cmdHandler} is executed as part of the setup. Our utility class \texttt{ServiceProc} extends \texttt{subprocess.Popen}, and is used to wrap the application execution. It performs two main tasks: (i) it internally caches the process \texttt{stdout} and \texttt{stderr} so that they can be examined later; and (ii) it can optionally block until the supplied \texttt{expect\_str} appears in \texttt{stdout}. In this case ``About to create listening address'' is written to \texttt{stdout} when the application has fully started and is ready to accept commands. This latter feature makes it possible to start our multi-process system in a predictable fashion. An additional feature of \texttt{ServiceProc} is that it will print whatever was captured from the subprocess \texttt{stdout} and \texttt{stderr} up to the point \texttt{expect\_str} was encountered. This output is in turn captured by \textsf{pytest}, and can be reported to the user with a label clearly indicating which part of the test setup produced it. Notably, if the \texttt{expect\_str} is {\em not} encountered within a timeout period (e.g., the process exited with an error prematurely), the \texttt{ServiceProc} constructor will print all of the encountered output of the process and throw an \texttt{Exception} which \textsf{pytest} will notice, and the reason for the failure will be immediately made obvious from the print statement (which will show any error messages, and indicate which fixture failed).

When a test that depends on this fixture concludes (either through normal execution, or if an \texttt{Exception} is raised), \textsf{pytest} will execute everything after the \texttt{yield} statement, for each of the fixtures, in the inverse order of startup. The \texttt{stopProc()} method of our \texttt{ServiceProc} class prints all of the captured \texttt{stdout} since \texttt{expect\_str} was encountered. Again, this output is captured by \textsf{pytest} and can be displayed if requested. This is particularly useful when a test fails: the \textsf{pytest} output can show the captured output from the setup and teardown of each fixture, in addition to the particular test assertion that failed.

We have written a native Python interface for the HEART telemetry streaming protocol to enable the transfer of information between the Python-based SRT application and C applications, but it is also useful for testing. The following is an example test fixture used to start up a global telemetry stream object called \texttt{hoPsdTelemetryStream} that will start a thread to asynchronously receive data from the running RTC, and which can later be queried by a test to see what was received. At teardown the connection is closed. It depends on the \texttt{cmdHandler} fixture shown above:

\footnotesize
\begin{Verbatim}[frame=single]
@pytest.fixture
def hoPsdStreamReader(cmdHandler):
    """
    Startup and shutdown of HO PSD telemetry stream reader
    """

    # === Setup ===

    global hoPsdTelemetryStream

    hoPsdTelemetryStream = TelemetryStream("", SRT_HO_PSD_PORT)
    hoPsdTelemetryStream.waitForConnection(5)
    print("HO PSD telemetry stream connected")

    yield hoPsdStreamReader

    # === Teardown ===
    
    if hoPsdTelemetryStream is not None:
            hoPsdTelemetryStream.close()    
\end{Verbatim}
\normalsize
    
Another common ingredient in our tests is the synchronous execution of a short-lived applications, such as running a WFS simulator to send a burst of pixels and then exit, or to send a command using our stand-alone command client. We have a helper routine called \texttt{run\_program} that wraps \texttt{subprocess.run} (a blocking call) to assist with the capture of short-lived process outputs. In the following example we instruct the RTC to start by issuing a \texttt{RUN} command using the command-line client and to verify that it was accepted. Note that in the context of HEART, \texttt{RUN} puts the RTC into a state where it begins to accept wavefront sensor pixels at the full framerate and performs reconstruction of the measured wavefront, but does not actually drive any of the wavefront correctors (i.e., the AO control ``loop'' is not yet closed). A subsequent \texttt{CORRECT} command would then enable correction.

\footnotesize
\begin{Verbatim}[frame=single]
    clntProc = run_program([clientPrg, "-cmdName", "RUN" ], timeout=clientTimeout)
    assert "cmd<RUN>, ack<0><ACCEPTED><>, status<0>" in clntProc.stdout
\end{Verbatim}
\normalsize

Putting these elements together, the following shows example snippets from an actual test, including assertions about RTC outputs:

\footnotesize
\begin{Verbatim}[frame=single]
def test_scaoTemplateValidateTests_pyrPix_modal(hoPsdStreamReader):
    """
    \test \ref test_scaoTemplateValidateTests_pyrPix_modal
    \brief Feed pipeline known WFS inputs and check correctness of results
    """
...
    # Instruct the RTC to start calculating HO PSDs
    clntProc = run_program([clientPrg,
                            "-configEnableHrtFlag", "1",
                            "-configEnableHrtFlagField", "enableHoPsd",
                            "-blockCmd", "4100",
                            "-port", "5036"],
                        timeout=clientTimeout)
    assert "cmd<8>, ack<0><><>, status<0>" in clntProc.stdout
...
    # Instruct the RTC to start accepting pixels
    clntProc = run_program([clientPrg, "-cmdName", "RUN" ], timeout=clientTimeout)
    assert "cmd<RUN>, ack<0><ACCEPTED><>, status<0>" in clntProc.stdout
...
    # Instruct the RTC to start correcting
    clntProc = run_program([clientPrg, "-cmdName", "CORRECT" ], timeout=clientTimeout)
    assert "cmd<CORRECT>, ack<0><ACCEPTED><>, status<0>" in clntProc.stdout
...
    # Run WFS simulator to send some pixels to the RTC
    wfsProc = run_WfsSimulator(wfsDataPath)
    assert wfsProc.returncode == 0    
...
    # Verify deformable mirror commands
    dmCmd, dmCmdFile = load_testData("test_cbDmCmd0.fits")
    for i in range(NUM_FRAMES):
        np.testing.assert_allclose(dmCmd[i, :].flatten(), dmCmdExpected,
                                   rtol=rtoldm,
                                   err_msg='dmCmd frame '+str(i))
...
        # Verify "buckets" of data received by hoPsdTelemetryStream
        bucket = hoPsdTelemetryStream.getBucket(timeout=1)
        np.testing.assert_allclose(bucket.data,
                                   psdExpected[:, :, i],
                                   rtol=psd_rtol,
                                   err_msg='PSDs bucket '+str(i))
\end{Verbatim}
\normalsize

The range of tests that we have currently implemented using these techniques has allowed us to perform basic validation of HEART-based RTCs running in various modes, i.e., the RTC produces expected outputs given known inputs. However, there are two shortcomings that we wish to address in the near future. First, many of our tests currently use hard-wired addresses and port numbers for socket communications. This makes it challenging to run multiple tests concurrently on a single server. Second, these tests do not typically run the RTC at realistic rates (e.g., $\sim$500--2000\,Hz). In order to ensure that the results are predictable and can pass on a wide range of system setups (including developer laptops with limited memory and CPU cores!), the rates and sizes of the data vectors passed through the system tend to be small. These are two major areas of focus for the new test system described in Section~\ref{sec:new}

\subsection{Graphical User Interfaces}

The HEART engineering graphical user interfaces (GUIs) are web-based. The webserver is written in C like most of HEART, and uses \texttt{libwebsockets} to interact with the browser. The front end is a mixture of Javascript and HTML. At the time of writing we are re-factoring our initial prototype GUI that was used to operate REVOLT into components that can more easily be organized into custom interfaces for other RTCs. While we do not yet have any automated tests of the browser interface, we have started to write stand-alone HTML pages that use Javascript functions to provide simulated data into individual widgets to demonstrate their functionality. Once this initial refactoring work is completed we may investigate browser automation tests (e.g., using \textsf{Selenium}\footnote{https://www.selenium.dev/}). Separately, we perform basic system integration testing with the web server, such as verifying that it successfully starts up and is able to receive telemetry streamed from the RTC.

%
%
\section{New build and test environment}
\label{sec:new}

Recently we have begun implementing new build and test infrastructure to address several issues that are alluded to throughout this paper. The following points summarize the main goals of this effort:

\begin{itemize}

    \item \textbf{Reduce number of tools:} At the moment our infrastructure uses self-hosted instances of GitLab-CE, Jira, and Jenkins. We are planning to drop the usage of Jenkins since GitLab is also very capable of executing tests. This will reduce some of our tool maintenance effort.

    \item \textbf{Simplify testing on development branches:} Our automated test suite is currently only triggered by commits/merges to the master branch. The new system will make it easy to execute tests on branches before merging, so tests should always pass on master.

    \item \textbf{Facilitate automated testing for multiple operating systems:} HEART needs to support RTCs running on different operating systems to meet observatory requirements.
    
    \item \textbf{Increase testing throughput by enabling parallel builds:} Our current build host can only accommodate a single execution of the test suite at a time. We wish to remove this bottleneck by enabling parallel build and test execution.

    \item \textbf{Execution performance testing:} Previously we have only executed performance tests on an {\em ad hoc} basis. We are currently optimizing the system configurations of several rack-mounted servers that can used to run more realistic tests on a regular basis.

\end{itemize}

To address many of these points, we have decided to move away from Jenkins for test execution, and will instead use GitLab CI/CD
\footnote{https://docs.gitlab.com/ee/ci/}. While Jenkins is highly configurable and has served us well, experience from colleagues at the Dominion Radio Astronomical Observatory (DRAO) has shown that GitLab is perfectly capable of the types of testing that we wish to perform, and would also reduce the number of tools that we need to maintain. The GitLab CI/CD system uses ``test runners'' (separate processes) to execute particular jobs when requested. A key part of our system is that we will execute all of our existing tests (unit tests, component, and system validation tests) using Docker containers. Finally, we are preparing to execute new performance tests on servers that have been specifically configured for this task.

\subsection{GitLab workflow using Docker containers}

\begin{figure}[hbt]
    \begin{center}
        \includegraphics[width=0.9\linewidth]{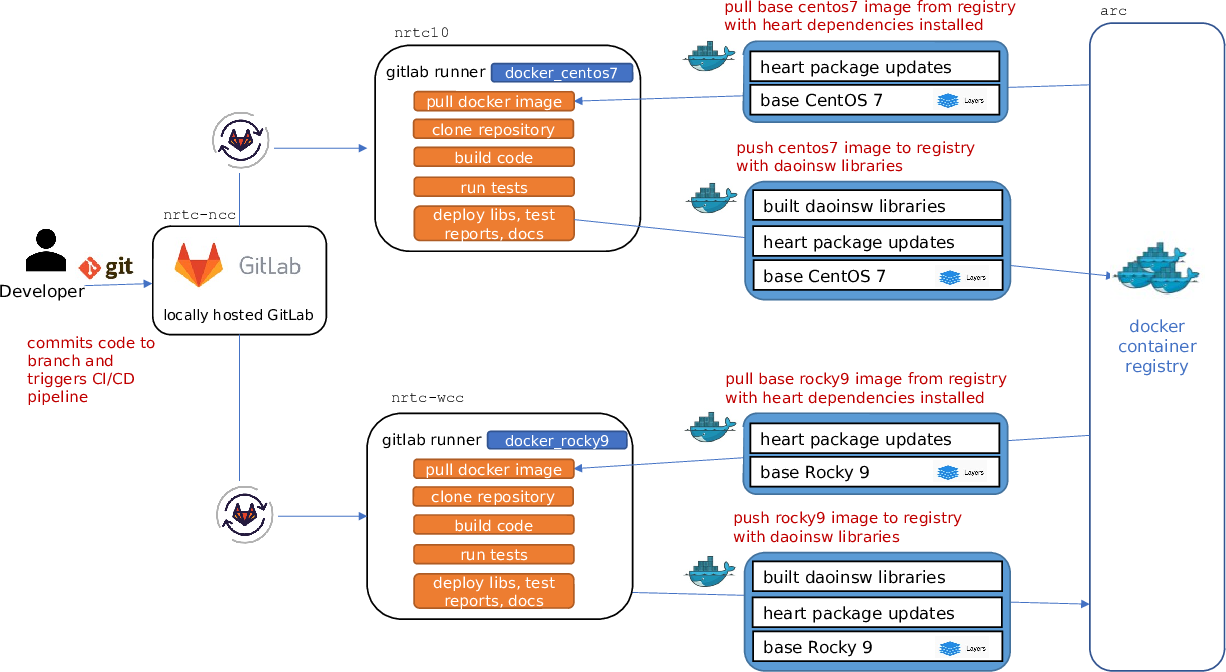}
    \end{center}  
    \caption{\label{fig:gitlab_flow}
    New GitLab workflow using Docker containers. As lower level dependencies such as DAOINSW are built they are stored in Docker images which may then be retrieved for use in subsequent jobs that depend on them. The central bubbles labeled \texttt{nrtc10} and \texttt{nrtc-wcc} are different physical servers on which the builds and tests will occur.
    }
\end{figure}

\begin{figure}[hbt]
    \begin{center}
        \includegraphics[width=0.9\linewidth]{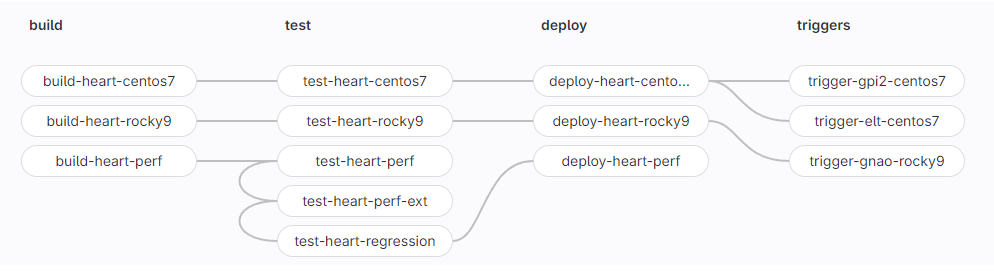}
    \end{center}  
    \caption{\label{fig:gitlab_pipeline}
    Example of how GitLab pipelines might be configured to produce builds and execute tests for specific RTCs running on different operating systems.
    }
\end{figure}

A Docker container shares the running kernel of the host operating system, but otherwise provides a local runtime environment, including a file system and a virtual network, making it easy to isolate code running in one container from another. This feature will immediately allow us to execute multiple test suites in parallel, since each container's virtual network will ensure that there are no socket communication conflicts (as noted at the end of Section~\ref{sec:hw}). Docker containers are instantiated from ``images'', whose file systems may consist of many layers. For HEART development we will have separate base images for all of the target operating systems that HEART must support (e.g., CentOS 7, Alma Linux 9, Ubuntu 22.04), including all of the core build dependencies (e.g., compilers, \textsf{Doxygen}, correct Python version). These images will be stored in our local Docker ``registry''. Any of our systems with the Docker daemon installed will then be able to retrieve these images from the registry as needed at runtime. When a GitLab runner is asked to build and test the lowest-level dependency in HEART, the DAOINSW library, it will start a Docker container from one of the base operating system images, pull the most recent commits on the branch being tested from the repository, compile the source code and run the tests. In addition to reporting back the results of the tests via the runner to GitLab, the build artifacts from DAOINSW will be persisted as a new layer in a Docker image stored in the registry. Then, when a downstream build and test job is required (e.g., HEART itself), this most recent DAOINSW image will be available as a starting point. Since each job simply obtains the most recent base image at the start of the job, new builds of upstream dependencies can commence while downstream builds are occurring. In other words, not only can multiple HEART builds jobs occur in parallel (because of the network isolation), but so can a new build of DAOINSW as it won't affect any of the already-running containers. An overview of this new workflow is shown in Figure~\ref{fig:gitlab_flow}, with examples of individual build and test pipelines in Figure~\ref{fig:gitlab_pipeline}.

\subsection{Execution performance tests}

Unlike all of the testing mentioned earlier, execution performance\footnote{We use the terms ``execution performance'' and ``performance'' interchangeably throughout this paper, not to be confused with ``AO performance'' (which would be related to things like the achieved Strehl ratio, which we do not cover here).} measurements are conducted under more stringent conditions. These new tests are being constructed to exercise HEART at scales and rates that are more representative of what will be expected for production systems.

For our CPU-based approach to RTCs, systems must be configured to minimize latency and jitter. First, we install the real-time kernel and whatever \texttt{rt-setup} package is provided for the Linux distribution that we are using. This creates the \texttt{realtime} group, and establishes limits associated with the group. We then use \texttt{tuna} to control various performance related Linux system parameters. We create a \texttt{tuned} directory for the RTC in, e.g., \texttt{/etc/tuned/heart}. The \texttt{tuned-adm} command is used to activate the profile. One of the most important uses of this system is to isolate (or restore previously isolated) CPU cores. We also modify parameters related to network latency such as busy reading (\texttt{net.core.busy\_read}) and buffer sizes (e.g., \texttt{net.core.rmem\_max}), and disable things like energy saving mode. We can further use a script in the profile directory to perform other \texttt{tuna} operations such as moving specific interrupts to a specified core when the RTC profile is enabled, or \texttt{ethtool} operations to configure the network adapter for lower latency performance. Note that we will have multiple profiles on a given system to support different types of tests (e.g., different AO modes, target RTC configurations). For example, the physical servers in Figure~\ref{fig:gitlab_pipeline} would have different profiles active depending on the tests they are running.

We are in the process of reconfiguring several servers with a dedicated 10\,Gb Ethernet switch for interconnects that will be used almost exclusively for this purpose. Unlike the containerized tests that we described earlier (which can run concurrently), performance tests can (generally) only be executed sequentially. We therefore envisage running these tests in two ways:

\begin{itemize}

\item \textbf{On demand:} A developer should be able to trigger a performance test manually through GitLab for a given branch. This will be useful for getting feedback during development to get a sense of the real-time execution properties of code after making changes.

\item \textbf{Nightly tests:} In order to measure the baseline performance of HEART over time, and potentially detect performance regressions, we will schedule performance tests every evening. We will store these results so that long-term trends can be measured.

\end{itemize}

How will execution performance be measured? Timing measurements are built into HEART in two principle ways. First, the circular buffer system includes a header for every ``bucket'' (our generic term for a frame of data) with time stamps that are filled by the writing process. Circular buffers can be recorded to disk using the telemetry system, and a separate analysis package, \texttt{heart-ana}, can load the files and match frames from different circular buffers to get individual end-to-end timings for the critical path through the HRT system. An example of AO system performance measurements using HEART telemetry is presented in another paper in these proceedings \cite{vankooten2024b}. Second, each HEART block records start and completion times to timestamp circular buffers that are read asynchronously by a performance monitor thread. All of the timings are measured with respect to a starting frame time that corresponds to the arrival of the first pixels in a designated WFS input block. The performance monitor then records the individual timings with respect to this reference in histograms that are periodically reset (e.g., every 10\,s). Each completed histogram is itself stored in a circular buffer that can be recorded to disk. This technique allows us to compress most of the pertinent timing information into a far more manageable size.

\begin{figure}[hbt]
    \begin{center}
        \includegraphics[width=0.75\linewidth]{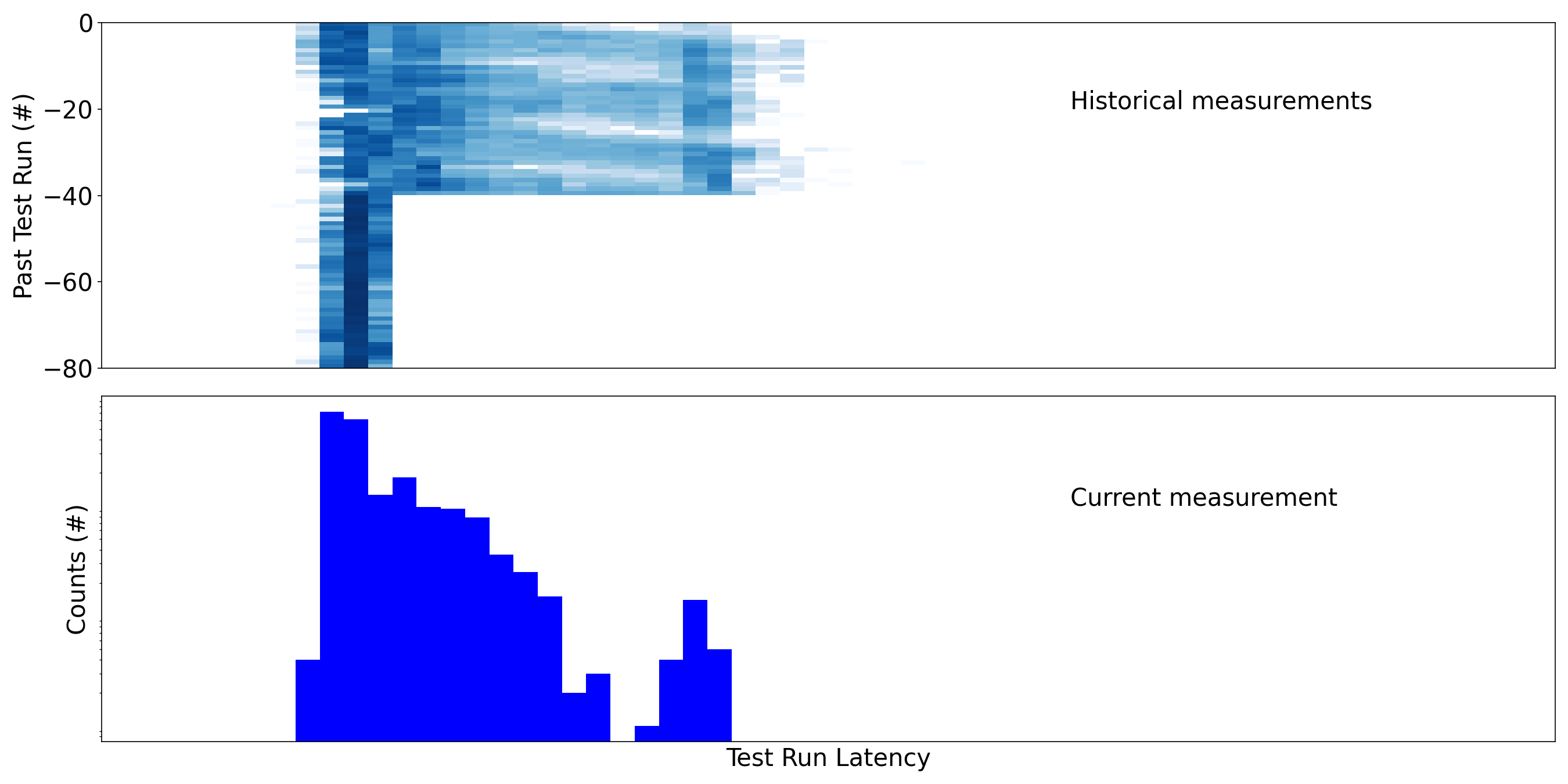}
    \end{center}  
    \caption{\label{fig:timing}
    Cartoon illustrating the type of information that can be recorded by nightly performance tests. The bottom plot is a histogram of timing latencies for a particular test run, exhibiting an anomalously large spread. The top window shows the historical trend of latencies over many days -- each row represents the histogram from a single night, and the intensities the heights of the bins. This representation makes it clear that some change occurred in either the software or the system configuration 40 test runs ago that introduced substantial jitter into the system.
    }
\end{figure}

One way we might use this information is illustrated in Figure~\ref{fig:timing}. Each time our nightly performance test is executed we can generate latency histograms for each of the processing blocks. We can also render the history of measurements as a 2-dimensional intensity map. These nightly reports will assist us in identifying any improvements and/or regressions introduced to the code base or system configuration. Note that these reports are {\em not} intended to demonstrate compliance with the requirements for any particular RTC. The larger systems that we are developing can require significantly more performant (and expensive) servers than we currently have on-hand (particularly for the ELTs). Rather, our goal is to measure relative {\em trends} in performance as we steadily build HEART-based RTCs, to identify parts of the code that can be improved, and to minimize regressions.

%
%

\section{Summary and future work}

The initial build phase of the Herzberg Extensible Adaptive optics Real-Time Toolkit (HEART) has been supported by an automated build and test system using GitLab and Jenkins that provides continuous and extensive feedback to the team in the form of documentation, test and coverage reports, and immediate messages on a Slack channel if there are any failures. This approach has helped ensure the quality of our code as it is built, but also confidence has we have taken on several new Real Time Controller (RTC) projects necessitating on-going refactoring and design effort. Low-level unit tests and some component tests of our C code use \textsf{cmocka}, while our Python code uses \textsf{pytest}. We also use \textsf{pytest} for our black-box and system-level tests that involve interactions between multiple running processes. It has proven to be convenient for managing the startup and execution of the separate pieces of software that are involved, and we have a native Python implementation of the HEART telemetry streaming protocol which makes it easy to capture and examine some of the outputs of running RTCs.

With the local AO testbed REVOLT currently operating with the first functioning HEART-based RTC, and with the completion of two facility-class RTCs due to be completed in the next $\sim$year (for GPI 2.0 and GNAO), we are now updating our infrastructure to make it easier for developers to test changes on branches before they are merged, to improve throughput, and to better allow us to tune and detect regressions in execution performance. Much of the improvement will result from a move to Docker container-based testing via GitLab CI/CD (replacing Jenkins and a single build host), which will allows us to build and execute tests for multiple operating systems in parallel (in contrast with our current system which can only execute builds and tests sequentially using a single host and operating system). The second portion of this effort is focused on configuring a set of dedicated servers that are optimized for real-time performance, and will be available to execute a suite of performance tests on-demand during the day, and also regular nightly tests to track performance over time.

\acknowledgments

We thank David Del Rizzo, Nicholas Bruce, and Dustin Lagoy at the Dominion Radio Astrophysical Observatory for helpful discussions regarding the use and configuration of GitLab CI/CD and Docker containers.

\bibliography{refs} 
\bibliographystyle{spiebib} 

\end{document}